\documentstyle[preprint,aps]{revtex}

\def\bscco{Bi$_2$Sr$_2$CaCu$_2$O$_{8+\delta}$}

\begin{document}

\tightenlines

\title{Tilt Modulus and Angle-Dependent Flux Lattice Melting in the Lowest
Landau Level Approximation}
\author{G. Mohler and D. Stroud}
\address{Department of Physics, Ohio State University, Columbus OH 43210}

\date{\today}

\maketitle  

\begin{abstract}

For a clean high-T$_c$ superconductor, 
we analyze the Lawrence-Doniach free energy in a tilted
magnetic field within the lowest Landau level (LLL) approximation.  The 
free energy maps onto that of a strictly $c$-axis field, but with a reduced
interlayer coupling.   We use this result to calculate the 
tilt modulus $C_{44}$ of a vortex lattice and 
vortex liquid.   The vortex contribution to 
$C_{44}$ can be expressed in terms of the squared $c$-axis Josephson 
plasmon frequency $\omega_{pl}^2$.   The transverse component
of the field has very little effect on the position of the melting curve.

\end{abstract}

\newpage

\section{Introduction}

This paper is concerned with the tilt modulus $C_{44}$ of the vortex
system in a high-T$_c$ or otherwise layered superconductor.
$C_{44}$ is an elastic constant which measures the free energy cost of
applying a small transverse field in addition 
to a field applied parallel to the $c$ axis.  It is a relevant parameter
in many physical processes, such as collective
pinning~\cite{larkin}, the ``peak effect,''\cite{larkin,peak}, and the
transition to the Bose glass state\cite{bose}.
Previous investigations have shown that $C_{44}$ is very strongly dependent
on the wave vector ${\bf k}$ of the transverse field\cite{sudbo,goldin}.  In
addition, it is finite in the flux liquid as well as the 
flux solid phase and is strongly affected by
different kinds of disorder\cite{larkin,tauber,benetatos}

A number of authors have analyzed $C_{44}$ in various approximations.  
Several groups\cite{sudbo,goldin,koshelev1}
have calculated $C_{44}({\bf k})$ in a flux lattice as a function of 
the wave vector ${\bf k}$ of the transverse magnetic field.  Other authors
have considered $C_{44}$ in the flux liquid state\cite{benetatos,larkin1}, and
in the presence of disorder in both the solid and liquid 
phases\cite{tauber,benetatos}.

In this paper, we calculate $C_{44}$ for a layered superconductor
at {\em high fields}.  We describe the superconductor using
a Lawrence-Doniach\cite{doniach} free energy functional,
and we evaluate fluctuations in the lowest Landau level (LLL) approximation, 
appropriate for strong $c$-axis magnetic fields.  
This LLL approach accounts well for the position of the flux lattice 
melting curve in the magnetic-field/temperature plane\cite{sasik1,hu1}, as 
well as the value of the magnetization in 
both solid and liquid phases\cite{sasik1,sasik2}, both in
\bscco and the less anisotropic
YBa$_2$Cu$_3$O$_{7-x}$.  To obtain $C_{44}$, we generalize the 
LLL approximation
to apply to fields with a finite $ab$ component.  
For a defect-free system in the LLL approximation, the 
tilted field free energy can be exactly mapped onto the usual free energy 
for a field ${\bf B} \| c$, but with a weaker interlayer 
coupling which depends on tilt angle.  

As a byproduct of this transformation, we can also analyze the behavior
of the flux lattice melting curve in a magnetic field tilted at an angle
to the $c$ axis.  We find that this curve is little affected by the presence
of a transverse magnetic field component at fields where the LLL approximation
is applicable, consistent with available (but limited) experiment.  

\section{Lowest Landau Level Approximation with Tilted Field}

At fixed external magnetic field ${\bf H} = H_z{\mathbf \hat{z}}
+ H_x{\mathbf \hat{x}}$
the Lawrence-Doniach Gibbs free energy in a form which 
incorporates the field energy is\cite{doniach}
\begin{eqnarray}
G[\psi, {\bf A}] &=& d\sum_n \int d^2 {\mathbf r} \,
\left\{ a(T) \left| \psi_n({\mathbf r}) \right|^2 
+ t \left| \psi_n({\mathbf r})e^{-i \frac{2\pi}{\Phi_0}A_zd}
 - \psi_{n-1}({\mathbf r}) \right|^2 \right.\nonumber \\
&& \left. +\, \frac{1}{2m_{ab}}
\left| \left( -i\hbar \nabla_{\perp} - \frac{q}{c}{\mathbf A}_{\perp} \right)
\psi_n({\mathbf r}) \right|^2
+ \frac{b}{2}\left| \psi_n({\mathbf r}) \right|^4 \right\} 
+ V\frac{|{\bf B} - {\bf H}|^2}{8\pi} \nonumber\\
& \equiv & \tilde{G}[\psi,{\bf A}] + V\frac{|{\bf B} - {\bf H}|^2}{8\pi}.
\label{eq:free}
\end{eqnarray}
Here $\psi_n$ is the order parameter in the $n^{th}$ layer,
$d$ is the distance between layers, $q = -2|e|$, and
$t = \hbar^2/(2m_cd^2)$ is the interlayer coupling energy.  In a cuprate
superconductor such as \bscco, $d$ represents the repeat distance 
in the $c$ direction.  ${\bf \nabla}_{\perp}$ is the $xy$ component of
the gradient operator, $V$ is the system volume, and the magnetic induction
${\bf B} = {\bf \nabla} \times {\bf A}$, where  
the vector potential ${\bf A} = ({\bf A}_{\perp}, A_z)$.

We choose the gauge ${\mathbf{A}} = -B_zy{\mathbf \hat{x}}
+ B_xy{\mathbf \hat{z}}$ and expand the $\psi_n$'s as a linear combination
of LLL states, to obtain
$\psi_n({\mathbf r}) = \left(\frac{\sqrt{3}a_H^2(T)}{4b^2}\right)^{1/4}
\sum_k c_{k,n} e^{ikx - (y-k\ell^2)^2/2\ell^2}$,
where $k = 2\pi p/L_x$ is the quantized momentum ($p$ a positive integer),
$\ell = \sqrt{\Phi_0/2 \pi B_z}$ is the magnetic length,
$a_H(T) = \left[a(T)+2t\right]\left[1-B_z/H_{c2}(T)\right]$, and
sample volume is $V = L_xL_yL_z$.  In terms of measurable quantities,
$a(T) +2t = -\hbar q H_{c2}(T)/(2m_{ab}c)$, where $m_{ab}$ is the
transverse effective mass and $H_{c2}(T)$ is the upper critical field as
a function of temperature $T$.  
Substituting this expansion into equation~(\ref{eq:free}) and 
integrating over $x$ and $y$ yields
\begin{eqnarray}
\tilde{G} & = &\frac{a^2_H(T)\pi \ell^2 d}{b} N_x\sum_{k,n}
\left\{ -|c_{k,n}|^2 - \frac{t\exp\left(-(\pi B_x d \ell / \Phi_0)^2\right)}{|a_H(T)|}
(c_{k,n}^*c_{k,n-1} e^{-i2\pi B_x d \ell^2 k/\Phi_0} + c.c.)\right. \nonumber \\
& + & \left.  \frac{3^{1/4}}{5^{5/2}} \sum_{q_1,q_2}
c_{k,n} c^*_{k+q_1,n} c^*_{k+q_2,n} c_{k+q_1+q_2,n}
e^{-(q_1^2 + q_2^2)\ell^2/2} \right\} 
\label{eq:free1}
\end{eqnarray}
where $N_x = (L_x/\ell)(\sqrt{3}/4\pi)^{1/2}$.  This expresses the Gibbs
free energy as a function of the expansion coefficients
$c_{k,n}$ for a tilted magnetic field.  This expression differs
from that for ${\bf B} \| \hat{{\mathbf z}}$\cite{sasik1} in only two respects: 
(i) there is an extra phase factor in the interlayer coupling; and (ii)
the strength of the interlayer coupling is renormalized by an exponentially
decaying factor.

The phase factor in eq.\ (\ref{eq:free1}) can be eliminated by introducing
a new set of coefficients
$\tilde{c}_{k,n} \equiv c_{k,n}e^{-i(2\pi B_x d \ell^2 k/\Phi_0)n}$,
in eq.\ (\ref{eq:free1}).  
The product term becomes
$c^*_{k,n}c_{k,n-1} e^{-i2\pi B_x d \ell^2 k/\Phi_0}
= \tilde{c}^*_{k,n}\tilde{c}_{k,n-1}$.
This transformation does not affect the
free energy term which contains products of four coefficients.
To see this, note that when a typical such product is transformed,
it picks up a phase factor
$e^{i(2\pi B_x d \ell^2 n/\Phi_0)(k-(k+q_1)-(k+q_2)+(k+q_1+q_2))}$, 
where each $(k+q_i)$ term is an integer modulo 
$N_\Phi$, $N_\Phi$ being the number of flux lines in the sample.
The terms in the exponential
can be summed to yield zero, leaving this term unaltered.
Similarly $|c_{k,n}|^2 = |\tilde{c}_{k,n}|^2$.  

In terms of the new variables, the free energy is thus
\begin{eqnarray}
\tilde{G} &=& \frac{a^2_H(T)\pi \ell^2 d N_x}{b} \sum_{k,n}
\left\{-|\tilde{c}_{k,n}|^2 -
\frac{t^\prime}{|a_H(T)|}
(\tilde{c}_{k,n}^*\tilde{c}_{k,n-1} + c.c)\right. \nonumber \\
& + & \left. \frac{3^{1/4}}{5^{5/2}} \sum_{q_1,q_2}
\tilde{c}_{k,n} \tilde{c}^*_{k+q_1,n}
\tilde{c}^*_{k+q_2,n} \tilde{c}_{k+q_1+q_2,n}
e^{-(q_1^2 + q_2^2)\ell^2/2}\right\},
\label{eq:newt}
\end{eqnarray}
where 
\begin{equation}
t^\prime = te^{-(\pi B_x d \ell / \Phi_0)^2}
\label{eq:tprime}
\end{equation} 
is the renormalized interlayer coupling.  Thus, the 
tilted B-field free energy is {\em identical in form} to the free energy
for a field ${\bf B} \| \hat{{\mathbf z}}$, but with a
{\em weaker} interlayer coupling.  
The ground state solution for the redefined amplitudes $\tilde{c}_{k,n}$
are identical to those for the $c_{k,n}$'s in a strictly
longitudinal field but weaker ${\bf B} \| \hat{{\mathbf z}}$.  However, 
the order parameter $\psi_n$ picks up a $B_x$ dependency,
\begin{eqnarray}
\psi_n({\mathbf r}) = \left(\frac{\sqrt{3}a_H^2(T)}{4b^2}\right)^{1/4}
\sum_k \tilde{c}_{k,n} e^{-ik(2\pi B_x d \ell^2 /\Phi_0)n}
e^{ikx - (y-k\ell^2)^2/2\ell^2},
\end{eqnarray}
corresponding, at low temperatures, to a tilted Abrikosov lattice.

\section{Tilt Modulus}

$C_{44}$ is defined by
\begin{eqnarray}
C_{44} = \frac{1}{V}
\left( \frac{\partial^2 {\mathcal G}}{\partial \theta^2} \right)_{\theta=0},
\end{eqnarray}
where $\theta$ is the angle between {\bf H} and the $c$ axis, and
where ${\mathcal G}$, the Gibbs free energy appropriate to an 
experiment at constant ${\bf H}$ and $T$, is given by 
\begin{eqnarray}
{\mathcal G} = -k_BT 
ln\int \prod_{k,n} d\tilde{c}^*_{k,n}\,d\tilde{c}_{k,n}\, e^{-\tilde{G}/k_BT}
+ V\frac{|{\bf B} - {\bf H}|^2}{8\pi} \equiv 
{\mathcal G}^v + V\frac{|{\bf B} - {\bf H}|^2}{8\pi}.
\label{eq:fe}
\end{eqnarray}
Then, writing $\partial/\partial\theta \rightarrow H_z\partial/\partial H_x$ for
small $H_x$, we find that the second term in eq.\ (\ref{eq:fe}) contributes
a term $C_{44}^c = H_z^2/(4\pi) \sim B_z^2/(4\pi)$ in the LLL
regime, where the magnetization is small.  This is the compressive
contribution to the tilt modulus.  

To calculate the remaining (vortex-related) contribution to $C_{44}$, 
which we denote $C_{44}^v$, we first write
$\tilde{G}/k_BT = {\mathcal H}/{\mathcal T}$, where
${\mathcal T} = \frac{bk_BT}{a^2_H(T)\pi \ell^2 d}$ and
\begin{eqnarray}
{\mathcal H} &=& N_x \sum_{k,n}
\left\{ -|c_{k,n}|^2 - \frac{t\exp\left(-(\pi B_x d \ell/
\Phi_0)^2\right)}{|a_H(T)|}
(\tilde{c}^*_{k,n} \tilde{c}_{k,n-1} + c.c.)\right.
\nonumber \\
& + & \left. \frac{3^{1/4}}{2^{5/2}} \sum_{q_1,q_2} \tilde{c}_{k,n} 
\tilde{c}^*_{k+q_1,n}
\tilde{c}^*_{k+q_2,n} \tilde{c}_{k+q_1+q_2,n} e^{-(q_1^2 + 
q_2^2)^2/\ell^2}\right\}.
\end{eqnarray}
Then writing
$\partial^2/\partial \theta^2 =B_z^2 \partial^2/ \partial B_x^2$, taking
$B_z \sim H_z$, and evaluating the derivatives from eqs.\ (\ref{eq:fe}), 
with the result
\begin{eqnarray}
C_{44}^v &=& \left(\frac{\sqrt{3}}{8}\right)^{1/2}
\left( \frac{\pi t d^3 |a(T)+2t|}{\sqrt{\Phi_0} b L_y L_z}\right)
(1-B_z/H_{c2})\sqrt{B_z}
\langle \sum_{k,n} (c^*_{k,n}c_{k,n-1} + c.c.)\rangle_T,
\label{eq:c44}
\end{eqnarray}
where we have dropped the tildes on the $c_{k,n}$'s because this expression
is evaluated at $B_x = 0$.
Eq.\ (\ref{eq:c44}) can be expressed in terms of measurable quantities
using the identities
$a(T) + 2t = \hbar q H_{c2}(T)/2m_{ab}c$ and
$b = 2\pi\kappa^2\left(q\hbar/m_{ab}c\right)^2$, where 
$\kappa = \lambda_{ab}(B_z, T)/\xi_{ab}(B_z, T)$ is the ratio of the  
$ab$ penetration depth and coherence length.
We also write $t = \hbar^2/(2m_{ab}d^2\gamma^2)$, 
where $\gamma^2$ is the anisotropy
parameter.  The result is
\begin{equation}
C_{44}^v = \frac{1}{32\pi\kappa^2\gamma^2}[H_{c2}(T)-B_z]B_z
\frac{N_x}{N_\Phi N_z}\langle \sum_{k,n}c_{k,n}^*c_{k,n-1} + c.c.\rangle_T.
\label{eq:c44a}
\end{equation}
Finally, the total tilt modulus is 
\begin{equation}
C_{44} \sim \frac{B_z^2}{4\pi} + C_{44}^v.
\label{eq:c44tot}
\end{equation}

In a triangular Abrikosov lattice, for example,
only $N_y = N_\phi/N_x$ of the $N_\Phi$ coefficients in each layer are
nonzero.  The nonzero coefficients are
$c_{2pN_x,n} = (2/\beta_A)^{1/2}$,  
$c_{(2p+1)N_x,n} = i(2/\beta_A)^{1/2}$,
where $p$ is an integer ranging from 0 to $N_y$ and 
$\beta_A = 1.159595...$ is the
Abrikosov ratio\cite{sasikthesis}.  
Substituting these values  into eq.\ (\ref{eq:c44a}), using 
$N_y/L_y = 2N_x/(L_x\sqrt{3})$, and adding the compressive term, we obtain
\begin{equation}
C_{44} = \frac{B_z^2}{4\pi}+
\frac{1}{8\pi\beta_A\kappa^2\gamma^2}(H_{c2}(T)-B_z)B_z.
\label{eq:abrik}
\end{equation}   

Eq.\ (\ref{eq:abrik}) can be compared to a previous 
estimate\cite{larkin1,gesh} 
$C_{44}^{LVG} = (B_z^2/(4\pi)
[1+1/(4\pi\tilde{\lambda}_{ab}^2(B_z, T)\gamma^2n_s)]$,
where $n_s$ is the superfluid number density for a 
two-dimensional bose system related to the
three-dimensional flux line system by a path integral
formalism\cite{bose}.  Writing $\lambda_{ab}(B_z, T) =
\kappa\xi_{ab}(B_z, T)$, 
$1/\xi_{ab}^2(B_z, T) \sim 2\pi(H_{c2}(T) - B_z)/\Phi_0$, and
approximating $n_s$ by $n_0 = B_z/\Phi_0$, the number of bosons 
(i. e., flux lines) per unit area\cite{tauber,benetatos}, we obtain
$C^{LVG}_{44} \sim B_z^2/(4\pi) + 
[H_{c2}(T)-B_z)B_z]/(8\pi\kappa^2\gamma^2)$
This result is in agreement with our own result,
eq.\ (\ref{eq:abrik}), except for the factor $(1/\beta_A) \sim 0.86$, which
arises from the nonuniformity of $n_s$ in the Abrikosov lattice phase. 

Our $C_{44}^v$ is closely 
connected to the so-called Josephson plasmon frequency $\omega_{pl}$, 
as calculated in the same LLL approximation\cite{hwang}.  
$\omega_{pl}^2(B_z,T)$ is the squared plasma frequency of the 
Josephson junction formed between adjacent $ab$ layers, and is given
by \cite{hwang}
\begin{eqnarray}
\omega^2_{pl}(B, T) = (\omega^{MF}_{pl})^2 
\frac{\beta_A}{2}[\beta_AN_x/(2N_{\Phi}N_z)]
\langle\sum_{k,n} (c^*_{k,n} c_{k,n-1} + cc)\rangle_T,
\label{eq:plasmon}
\end{eqnarray}
where $\omega^{MF}_{pl} =
\sqrt{[(H_{c2}(T)-B)cq]/[\epsilon_0\kappa^2\gamma^2\hbar\beta_A]}$
is the mean-field Josephson plasmon frequency 
and $\epsilon_0$ is an interlayer dielectric constant.
Combining eqs.\ (\ref{eq:c44a}) and (\ref{eq:plasmon}) gives
\begin{equation}
\frac{C_{44}^v(B_z, T)}{\epsilon_0\omega_{pl}^2(B_z, T)} = 
\frac{B_z \hbar^2}{8 \Phi_0 q^2}.
\label{eq:finaltilt}
\end{equation}

Fig.\ 1 shows $C_{44}^v(B_z,T)$ for $B_z = 2$ tesla
in clean \bscco, using this relation
$\omega_{pl}^2$ as calculated in Ref.\ \cite{hwang}.  
Like $\omega_{pl}^2$, $C_{44}^v$ 
has a discontinuity at the flux lattice melting transition, and remains
finite in the flux liquid state.
Experiments on $\omega_{pl}^2$ are consistent with this result, indicating
that $\omega_{pl}^2$, and hence $C_{44}^v$,  have discontinuities
at flux lattice melting in clean \bscco\cite{shibauchi1,matsuda}.

\section{Flux Lattice Melting in a Tilted Magnetic Field}

Eqs.\ (\ref{eq:newt}) and (\ref{eq:tprime}) have some striking implications
for the LLL phase diagram of a clean layered superconductor. 
When ${\bf B} \| c$, this phase diagram depends on only two
parameters, namely $g = a_H\sqrt{\pi\ell^2d/(bk_BT)}$ and 
$\eta = t/|a_H|$\cite{hu1}.
Our results show that, even with a transverse field $B_x$,
the phase diagram still depends on only {\em two} parameters, 
except that $\eta$ is now replaced by 
$\eta^{\prime} = \eta\exp[-(\pi B_x^2d^2/(2B_z\Phi_0)]$.  As found 
previously\cite{sasik1,hu1,sasik2},
this phase diagram contains a single first-order melting line separating 
a triangular vortex lattice from a vortex liquid.  In the regime where 
the LLL approximation is adequate, our results show that 
this first-order line should persist in a tilted magnetic field (cf.\ Fig. 2).  
Furthermore, in most high-T$_c$ materials, the line is shifted very little
by a nonzero $B_x$.  (For \bscco,  
$t^\prime\sim t\exp[-7.5\times 10^{-7}B_x^2/B_z]$.)

We have found no experimental melting data in tilted
magnetic fields at fields where the LLL approximation is applicable.
In BiSr$_2$Ca$_2$Cu$_2$O$_{8+x}$, the low-field melting temperature has been
reported\cite{schmidt} to depend only on the $c$ component of ${\bf H}$,
which, though obtained in a very different regime, would be consistent
with the result of our calculations. It would be of great 
interest to have further tests of the LLL predictions in the relevant
high-field regime.

\section{Summary}

In this paper, we have extended the LLL approximation
for high-T$_c$ superconductors to treat fields tilted at
an angle to the layer perpendicular.  The resulting
free energy has exactly the same {\em form} as the usual case, except
that the effective interlayer coupling is reduced.  For high-T$_c$ materials,
this reduction is small; hence, we predict that the flux lattice melting
temperature will be little affected by the application of an oblique
magnetic field in the range where the LLL approximation is
valid, except possibly at angles nearly parallel to the layers.  This
prediction appears consistent with some existing (but low-field)
experiments\cite{schmidt}.   We also obtain an expression for the 
zero-wave-vector tilt modulus $C_{44}$, in good agreement with previous
estimates by other means\cite{larkin1,gesh}.
Finally, the vortex contribution $C_{44}^v$ is proportional to the 
squared Josephson plasmon frequency $\omega_{pl}^2$, as calculated in the 
same LLL approximation, and remains finite in the vortex liquid as well
as the vortex solid phase.

\section{Acknowledgments}

This work was supported by the Midwest Superconductivity Consortium
through Purdue University, Grant No. DE-FG 02-90 ER45427, and  
by NSF grant DMR97-31511.

\newpage

\begin{center}
FIGURES
\end{center}
\vspace{0.1in}

\noindent
FIG.\ 1.  Vortex contribution $C_{44}^v$ to the tilt modulus,
plotted versus temperature for \bscco at $B_z = 2T$, as evaluated using eq.\ 
(\ref{eq:finaltilt}) and $\omega_{pl}^2$ from Ref.\ (\cite{hwang}).
$C_{44}^v$ is discontinuous at the flux lattice melting temperature $T_M(B_z)$
(indicated by arrow). Inset: enlargement of $C_{44}^v$ near and above
$T_M$.

\vspace{0.1in}

\noindent
FIG.\ 2.  Phase diagram of a clean high-T$_c$ material in the
LLL approximation,
including a transverse magnetic field component $B_x$.  The parameters 
$\eta^{\prime}$ and $g$ are defined in the text.   Points and spline fit
are given by Hu and MacDonald~\cite{hu1}.

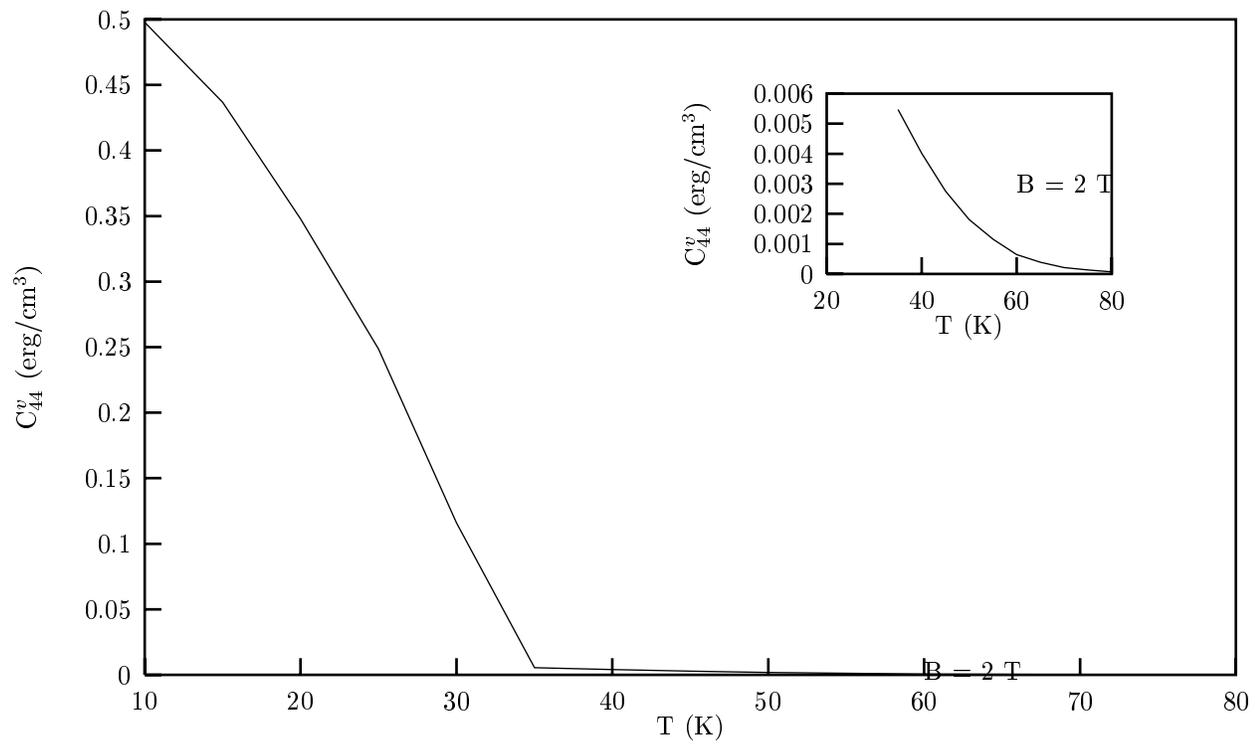
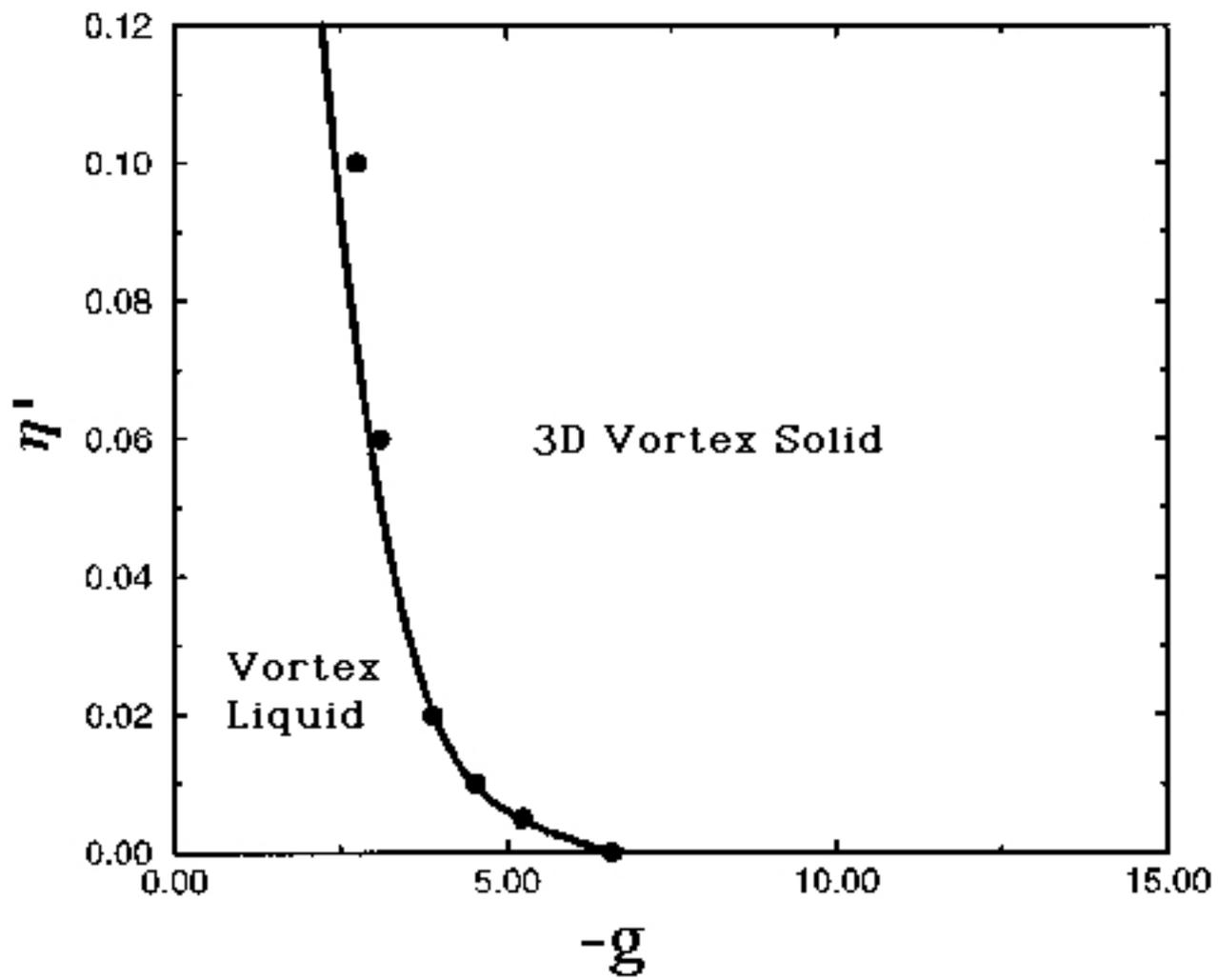
\end{document}